\documentclass[11pt]{article}
\pdfoutput=1
\usepackage{jcappub}
\usepackage{comment}
\usepackage{float}
\usepackage{amsfonts,amsmath,amssymb}
\usepackage{caption}
\usepackage{subcaption}
\usepackage{paralist}
\usepackage[normalem]{ulem}

\newcommand{\order}[1]{\ensuremath{\mathcal{O}(#1)}}

\newcommand{\stau}{{\ensuremath{\tilde{\tau}}}}
\newcommand{\sqrk}{{\ensuremath{\tilde{q}}}}

\newcommand{\msq}{{\ensuremath{m_{\sqrk}}}}
\newcommand{\diff}{\ensuremath{\mathrm{d}}}
\newcommand{\ldec}{\ensuremath{\lambda_\text{dec}}}

\newcommand{\psurv}{\ensuremath{p_\text{surv}}}
\newcommand{\elab}{\ensuremath{E_\text{lab}}}
\newcommand{\figref}[1]{\ref{fig:#1}}
\newcommand{\Gen}{IceCube-Gen\textit{2}}
\newcommand{\thzen}{\ensuremath{\theta_\text{Zenith}}}

\begin{document}

\title{Beyond standard model particles in the atmospheric flux: a long-lived stau  example}

\author[a]{Atri Bhattacharya,}
\author[b]{Mary Hall Reno,}
\author[c]{Ina Sarcevic}

\affiliation[a]{STAR Institute, University of Liège, Liège 4000, Belgium}
\affiliation[b]{Department of Physics and Astronomy, University of Iowa, Iowa City, IA 52242, USA}
\affiliation[c]{Department of Physics and Department of Astronomy,
	University of Arizona, Tucson, Arizona 85721, USA}

\emailAdd{A.Bhattacharya@uliege.be}
\emailAdd{mary-hall-reno@uiowa.edu}
\emailAdd{ina@physics.arizona.edu}

\date{\today}
\abstract{
	We discuss fluxes of long-lived supersymmetric (SUSY) particles produced in
	the atmosphere from ultra-high energy cosmic-ray interactions with air nuclei. We consider long-lived particle production which 
	 proceeds first via the on-shell formation of heavier particles in the
	SUSY spectrum and then their decays.
	Specifically, we focus on the production of stau pairs from the decays of
	squarks produced in the initial cosmic ray-air collision under the purview of
	gauge-mediated symmetry breaking models that have the stau as the
	next-to-lightest SUSY particle.
	The calculation of the resulting stau flux schematically mirrors that of prompt
	atmospheric neutrino production, however, with the energy scale of initial hadronic 
	collision set to EeV energy scales of the incident cosmic rays rather than the PeV
	energies where the prompt atmospheric neutrino fluxes dominate over neutrino fluxes from pions and kaons, and with the appropriate stau production cross section.   
	We show the effect of accounting for the energy distribution of the
	staus relative to the squark energy distributions.  We discuss the potential for
	future ultra-high energy detectors similar in scope to \Gen\ to detect
	atmospheric staus via tracks going through the detector volume. Current collider constraints on stau pair production within the SUSY framework considered here are more constraining than those from a detector like \Gen\ over ten years. 
}

\maketitle

\flushbottom

\date{\today}

\section{Introduction}
\label{sec:intro}

Cosmic ray--air interactions in the atmosphere produce air showers and yield fluxes of neutrinos
\cite{Barr:2004br,Honda:2006qj,Sinegovskaya:2014pia,Gaisser:2002jj,Lipari:1993hd,Pasquali:1998xf,Enberg:2008te,Gauld:2015kvh,Bhattacharya:2015jpa,Bhattacharya:2016jce,Zenaiev:2019ktw,Fedynitch:2022vty} that are detected in underground experiments \cite{Super-Kamiokande:2015qek,IceCube:2015mgt,IceCube:2017cyo,ANTARES:2021cwc,Kochanov:2021hkj}.
Cosmic ray energies per nucleon of $E_{\rm CR}\sim 10^8$~GeV incident on air nuclei have centre-of-mass energies $\sqrt{s}$ comparable to centre-of-mass energies of
$pp$ collisions at the Large Hadron Collider (LHC). Cosmic ray energies up to $\sim 10^{11}$~GeV have been measured in air shower experiments \cite{TelescopeArray:2018bya,PierreAuger:2020qqz}, pushing $\sqrt{s}$ beyond the LHC and opening opportunities to probe physics beyond the standard model (BSM) in energy regimes higher than accessible with man-made accelerators.

The potential to use cosmic rays to probe BSM physics has been explored in several contexts, for example, in cosmic ray and ultrahigh-energy neutrino interactions  in the atmosphere or in the Earth that produce long-lived quasi-stable supersymmetric (SUSY) charged particles such as staus ($\stau$), the supersymmetric partner of the tau \cite{Albuquerque:2003mi,Albuquerque:2006am,Reno:2005si,Huang:2006ie,Ahlers:2006pf,Ahlers:2007js,Ando:2007ds,Meighen-Berger:2020eun}.
Once produced, for $m_\stau\gg m_\tau$,
staus have much reduced electromagnetic energy loss in propagating through the Earth as compared to muons and taus. As a result, high energy quasi-stable staus can, in principle,  
produce high energy signals in underground detectors like IceCube.

In a category of minimal models of \stau\ production from cosmic ray interactions in the atmosphere, the atmospheric \stau\ flux comes from a two-step process analogous to the prompt atmospheric neutrino flux.
The dominant source of \stau\ production by cosmic ray nucleon ($N$) interactions with atmospheric nuclei is through squark (\sqrk ) pair production and decay, with a cross section that is many orders of magnitude larger than for direct \stau\ pair production \cite{Ahlers:2007js}.
Schematically, an atmospheric \stau\ flux is generated by prompt \sqrk\ decays via $N\to \tilde{q}\to$\stau . The prompt atmospheric neutrino flux comes from cosmic ray interactions with air to produce charm particles that rapidly decay to neutrinos, denoted by $N\to c\to \nu$.
Semi-analytic solutions of the coupled cascade equations that determine the prompt atmospheric neutrino flux using $Z$-moments are well studied \cite{Lipari:1993hd,Pasquali:1998xf,Enberg:2008te,Gauld:2015kvh,Bhattacharya:2015jpa,Bhattacharya:2016jce,Zenaiev:2019ktw}.
Here, we use the same techniques to illustrate how this same $Z$-moment formalism can be applied to the atmospheric \stau\ flux from prompt \sqrk\ decays. An important feature of the $Z$-moment method is that it accounts for how the \sqrk\ and \stau\ energies are related to the incident cosmic ray nucleons energies that have a steeply-falling spectrum.

We use cosmic ray production of \stau\ via $N\to \tilde{q}\to$\stau\  to illustrate the method to determine the prompt atmospheric \stau\ flux.
In the next section, we outline the SUSY model used as a starting point for our example of \stau\ production. Section \ref{sec:zmoments} shows the evaluation of the prompt atmospheric \stau\ flux. For signals in detectors, evaluations of \stau\ propagation in the Earth and energy loss in the detector are required. In Section \ref{sec:stau-prop}, we briefly review \stau\ propagation that includes electromagnetic energy loss and the probability for decay. We then discuss the potential \stau\ signals in IceCube and \Gen -like detectors  given the atmospheric \stau\ flux introduced in Section \ref{sec:zmoments} for benchmark particle masses. 

This example is interesting in the context of IceCube-like detectors because \stau\ particles have the potential to penetrate the Earth and  arrive from relatively larger zenith angles than $\nu_\mu$ produced muons. Electromagnetic energy loss in the Earth of a heavy \stau\  is suppressed by a factor of $\sim m_\mu/m_\stau$ relative to muon energy loss. 
Using benchmark choices for supersymmetry particle masses, we conclude in Section \ref{sec:conclusions} that underground detectors like IceCube cannot constrain \stau\ masses in parameter spaces already constrained by LHC experiments \cite{CMS:2012wcg,ATLAS:2012urj,ATLAS:2014fka,ATLAS:2019gqq,CMS:2016kce} (see, however,  ref. \cite{Meighen-Berger:2020eun}).

\section{SUSY models with long-lived staus}
\label{sec:susy}

We consider a SUSY scenario with the \stau\ as the next-to-lightest supersymmetric particle (NLSP) as it fulfils two key features benefitting its detectability at
large neutrino telescopes like IceCube, \textit{viz.}:
\begin{itemize}
	\item it is a secondary particle from the prompt decay
	      of the primary particle directly produced in the cosmic ray-air collision;
	      and
	\item it is quasi-stable and does not have strong interactions.
\end{itemize}

Within the realm of minimal supersymmetry (MSSM), models where the
symmetry is broken via gauge-mediated interactions called gauge-mediated
supersymmetry breaking (GMSB) \cite{Giudice:1998bp} naturally lead to \stau s\
for a wide range of symmetry breaking scales.
In these models, the soft SUSY breaking terms are generated at the messenger
scale $M$ unrelated to the flavour scale $\Lambda_F$.
When the messenger sector consists of three or more generations, the resulting
SUSY spectrum includes the goldstino as the LSP and the \stau\ as the NLSP
for the breaking scale around $10$ TeV or higher.
For example, the benchmark SPS 7 from SUSY Snowmass Points and
Slopes~\cite{Allanach:2002nj} outlines a GMSB scenario where the NLSP is a
\stau\ with its mass in the range of a few hundred GeV.
One may then choose parameters such that the NLSP decays to the LSP promptly or
alternatively, for the NLSP to be quasi-stable.
The latter choice, with $c\tau \approx 10$ m, is interesting for neutrino
telescopes since high energy \stau\ tracks may persist for distances of up to a
few 1000 km in rock.
For these quasi-stable \stau\ scenarios, it is possible to detect staus via the
loss of energy across their tracks in an IceCube-like detector.

SUSY models with long-lived \stau\ are constrained by null results from searches
at colliders.
In particular, model independent bounds on the \stau\ mass come from LHC limits
on electro-weak Drell-Yan production of \stau\
pairs~\cite{CMS:2012wcg,ATLAS:2012urj,ATLAS:2014fka}, yielding $m_\stau
	\geqslant 289$ GeV~\cite{ATLAS:2014fka}.
More specifically with respect to GMSB scenarios (SPS 7) --- the most
convenient of all SUSY breaking scenarios to produce \stau\ as the NLSP,
searches for \stau\ pair production made at $\surd s=13$ TeV at both
ATLAS~\cite{ATLAS:2019gqq} and CMS~\cite{CMS:2016kce} constrain
$m_{\stau}\geqslant 430$--$490$ GeV.
These limits come primarily from production of \stau\ pairs, so they are
relatively, but not completely, model independent in the GMSB framework. The limits~\cite{ATLAS:2019gqq,CMS:2016kce} come
from studies of ionization energy losses as the long-lived particle traverses
the detector and from time-of-flight
measurements~\cite{Heisig:2011dr,Feng:2015wqa}.
The dependence of the model on the messenger scale for these energies is rather
weak, so these limits should be considered as fairly robust.
With $m_{\tilde{\tau}}=490$ GeV, the corresponding \sqrk\ mass is
$m_{\tilde{q}}=1.3$ TeV in this GMSB scenario.

Without specifying a full MSSM model, we choose to work with a generic GMSB
scenario with a breaking scale $\Lambda \approx 100$ TeV and three generations
of messenger particles.
We consider one scenario with $m_\stau = 290$~GeV, just
above model independent bounds on $m_\stau$. This mass is in tension with
GMSB-model specific searches at colliders, however, the lighter mass scale of the
model lends itself to improved detectability at
IceCube as described in ref.~\cite{Meighen-Berger:2020eun}. We take the corresponding squark mass to be $m_\sqrk = 415$ GeV guided by the GMSB SPS 7 framework.
We also consider the case where $m_\stau = 490$~GeV and $m_\sqrk = 1.3$ TeV.
This allows us to study models with long-lived staus which are not already
disfavoured by collider searches, as noted above.

\begin{figure}[htb]
	\centering
	\includegraphics[width=0.45\textwidth]{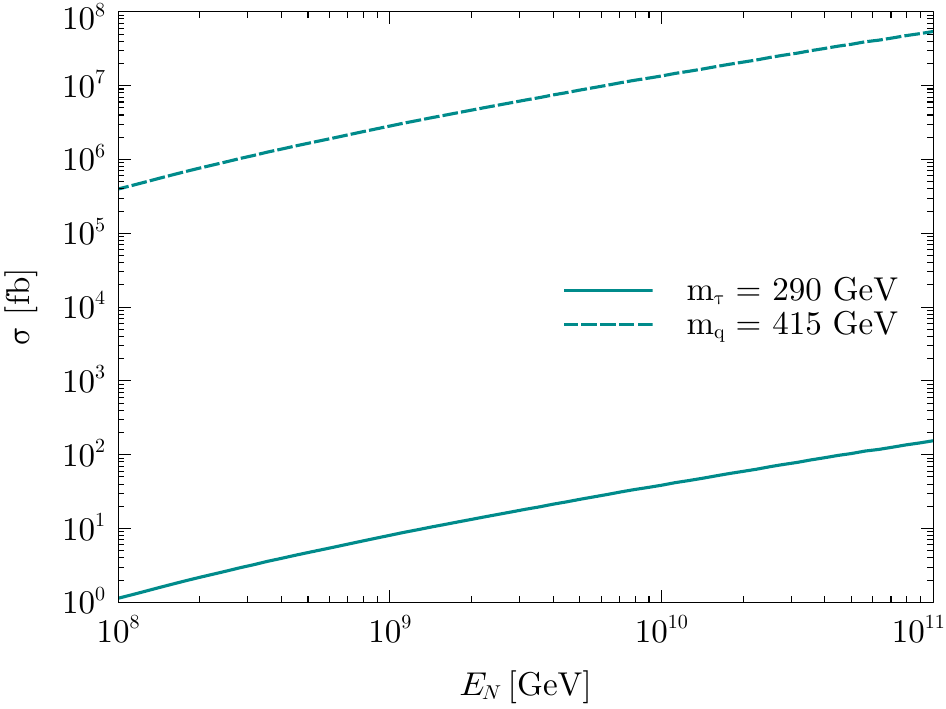}
	\includegraphics[width=0.45\textwidth]{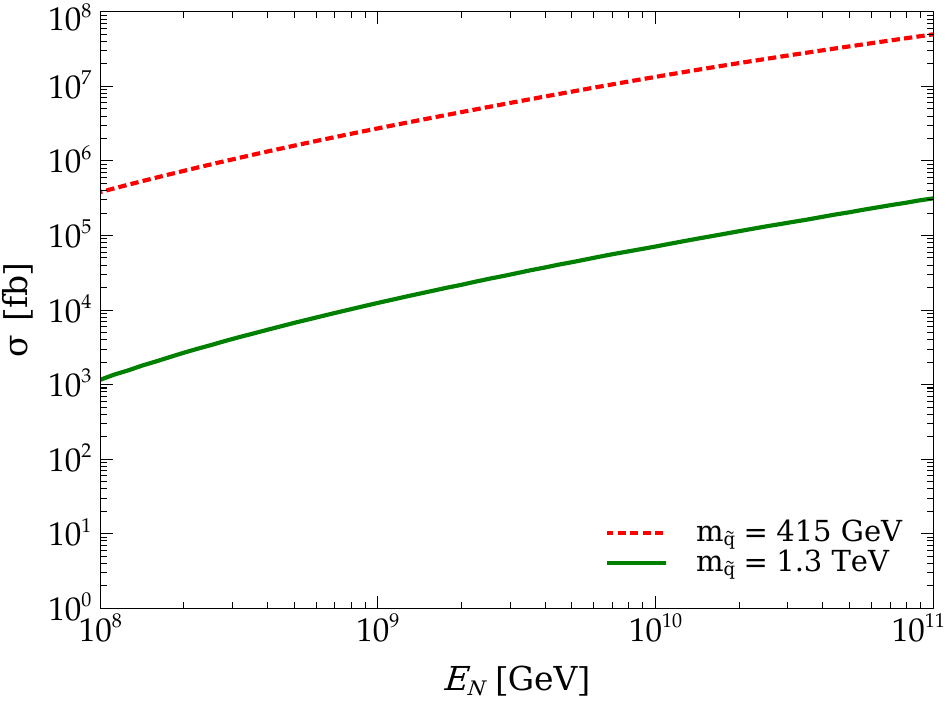} \\
	\includegraphics[width=0.45\textwidth]{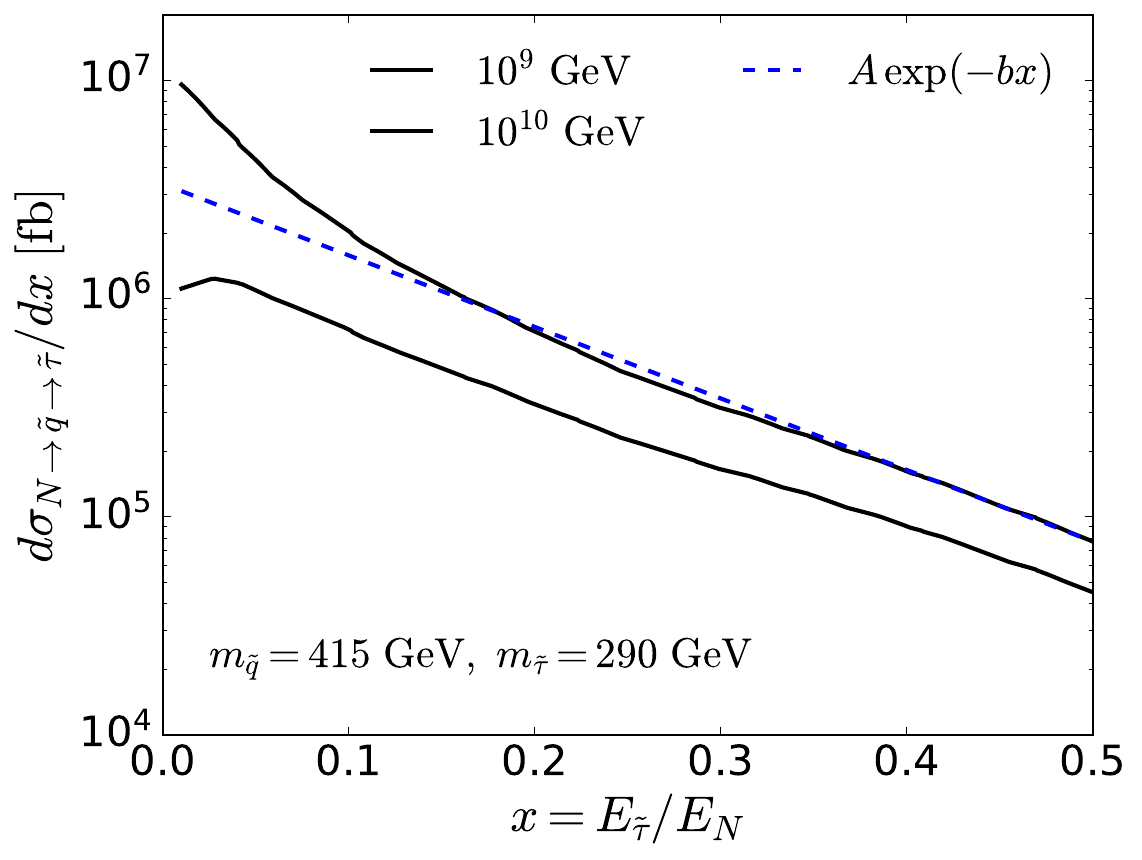}
	\caption{\label{fig:stau-xsec}{Top Left:} Direct $\stau$ pair production
		cross-section from $pp$ collisions in the atmosphere in GMSB scenarios, for $\stau$ pair production with 
		$m_\stau = 290$~GeV (solid) and $\sqrk$ pair production with $m_\sqrk=415$ GeV (dashed), as a function of the proton energy denoted $E_N$ in the fixed target frame. We assume a 100\%\ branching fraction for $\sqrk\to\stau$.
		The model independent collider limit on a long-lived stau is $m_\stau
			\geqslant 289$ GeV.
		{Top Right:} The cross-section for squark pairs with $\msq=415$ GeV (upper dashed curve)
		and $\msq=1.3$~TeV (lower solid curve).
		{Bottom:} The differential energy distribution of the lab frame $E_{\stau}$
		scaled by the incident cosmic ray nucleon energy for $E_N=10^9$ GeV (lower) and $10^{10}$ GeV (upper) for
		$\msq = 415$ GeV and $m_\stau=290$ GeV.
		The dashed line shows eq.~\eqref{eq:dsdx}.
	}
\end{figure}

In the GMSB SUSY model with quasi-stable staus, collisions between cosmic-ray
(CR) protons and atmospheric nuclei at the energies of $\elab \geqslant 1 $ EeV
lead to the production of staus by two distinct routes:
\begin{inparaenum}[\itshape a\upshape)]
	\item directly via the channel $pp \to \stau \bar{\stau} X$ and
	\item via the production of an intermediate squark that subsequently
	decays to produce $\stau$.
\end{inparaenum}
The cross sections for stau and squark production are shown in figure
\figref{stau-xsec}.
In the upper left panel of figure~\ref{fig:stau-xsec}, we show direct $\stau$ pair
production in the GMSB scenario for $m_\stau=290$ GeV (lower solid line). The
upper dashed curve shows the cross section of squark pair production with
$m_{\tilde{q}}=415$ GeV. The upper right panel of
figure~\ref{fig:stau-xsec} also shows the \sqrk\ pair production cross section for
$m_{\tilde{q}}=1.3$ TeV (lower curve).
Prompt decays of squarks will yield staus. We
assume a 100\%\ branching fraction for direct \sqrk$\to$\stau \ decay.

The cross section for direct \stau\ pair
production, even when assuming masses as low as $100$ GeV, is not high enough
to generate sufficient \stau\ fluxes from CR collisions in the atmosphere to enable \stau\
detection over and above the backgrounds at ultra-high energy detectors such as
IceCube and \Gen. The backgrounds come from a combination of atmospheric
and astrophysical muon neutrino fluxes that produce muons that traverse the detector and dominate the event rate from direct \stau\ pair
production in the atmosphere.
We therefore assess the dominant contribution to the \stau\ flux that comes
via the prompt decay of heavier particles in the spectrum, including squarks,
also produced in CR collisions with atmospheric nuclei.
This can increase the net \stau\ production cross section by 5--6
orders of magnitude, depending on the choice of squark and gluino
masses~\cite{Ahlers:2007js}.
In both \sqrk\ mass cases shown in the upper right panel of figure \ref{fig:stau-xsec}, direct stau production contributes a small fraction of the
total \stau\ flux, so we neglect it in our evaluations.

An important component in the evaluation of the \stau\ flux is the \stau\
energy distribution from the production, then prompt decay, of squarks. The
\stau\ energy distributions that result from the decays of these squarks,
scaled to the incident proton energy in fixed target collisions
($x=E_\stau/E_N$) for $\msq = 415$ GeV and $m_\stau=290$ GeV and for two
incident cosmic ray nucleon energies, $E_N=10^9$ GeV and $E_N=10^{10}$ GeV, is
shown in the lower panel of figure~\ref{fig:stau-xsec}.
The $x$ distributions for other energies and other masses have $x$
distributions that have approximately the same shape with different normalizations governed by 
$\sigma_{N\to\sqrk\to\stau}(E_N)$.
In what follows, we abbreviate $N\to\sqrk\to\stau$ by $N\to\stau$ and
approximate
\begin{equation}
\label{eq:dsdx-full}
	\frac{1}{\sigma_{N\to\stau}(E_N)}
	E_N\frac{d\sigma_{N\to\stau}(E_\stau,E_N)}{dE_\stau}= \frac{1}{\sigma_{N\to\stau}(E_N)}\frac{d\sigma_{N\to \stau} (x,E_N)}{dx}\simeq    \frac{dn_{N\to\stau}(x)}{dx}\,.
\end{equation}
The dashed line shows an analytic approximation to the $x$ distribution:
\begin{equation}
	\frac{d\sigma_{N\to \stau} (x)}{dx}\simeq A e^{-bx}
	\label{eq:dsdx}
\end{equation}
for $b=7.55$. We use this analytic approximation to the \stau\ energy distribution relative to the incident cosmic ray nucleon energy, and the approximate high energy behaviour of the cross section $\sigma_{N\to\stau}\propto E_N^{0.58}$ which is extracted from the upper right panel of figure 1,  as a cross check of our numerical results.

\section{The flux of prompt atmospheric staus}
\label{sec:zmoments}

\subsection{\texorpdfstring{$Z$}{Z}-moment formalism}

The determination of the atmospheric BSM particle flux incident on the Earth
follows from coupled cascade equations, analogous to the standard model
atmospheric lepton fluxes (see, e.g., Ref.~\cite{Lipari:1993hd}).
Generically, for particle species $i$ and $j$, the flux of particle $i$ $\phi_i(E,X)$ can be written as a function of atmospheric column depth $X$ according to
\begin{equation}
	\label{eq:master}
	\frac{\partial \phi_i(E,X)}{\partial X}=-\frac{\phi_i(E,X)}{\lambda_i(E)}
	-\frac{\phi_i(E,X)}{\lambda_i^{\rm dec} (E)}
	+\sum_j S_{j\to i}(E,X)\,,
\end{equation}
in terms of interaction length $\lambda_i(E)$, decay length
$\lambda_i^{\rm dec} (E)$, and source terms  $S_{j\to i}(E,X)$ that come from either production or decay. The source term for production can be approximately written as
\begin{eqnarray}
	\nonumber
	S_{i\to j}(E,X) &=& \int_E^\infty
	dE'\, \frac{\phi_i(E',X)}{\lambda_i(E')} \frac{dn_{i\to j}(E,E')}{ dE}\\ \nonumber
	&\simeq & \frac{\phi_i(E,X)}{\lambda_i(E)}\,\int_E^\infty\,dE'\,
	\frac{\phi_i(E',X=0)}{\phi_i(E,X=0)}\,
	\frac{\lambda_i(E)}{\lambda_i(E')} \frac{dn_{i\to j}(E,E')}{ dE}\\
	&\equiv & \frac{\phi_i(E,X)}{\lambda_i(E)}\, Z_{ij}(E)\,.
	\label{eq:zmom-def}
\end{eqnarray}
In eq.~\eqref{eq:zmom-def}, the distribution ${dn_{i\to j}(E,E')}/{dE}$ is
\begin{equation}
	\frac{dn_{i\to j}(E,E')}{dE} = \frac{1}{\sigma_i(E')}
	\frac{d\sigma_{i\to j}(E,E')}{ dE}\,,
\end{equation}
with incident particle $i$  energy $E'$ and outgoing particle $j$ of energy $E$.
The flux of cosmic ray nucleons ($i=N$, $j=N$) can be determined by
\begin{equation}
	\label{eq:nucleonflux}
	\frac{\partial \phi_N(E,X)}{\partial X}=-\frac{\phi_N(E,X)}{\lambda_N(E)}
	+S_{N\to N}(E,X)\,.
\end{equation}
The top of the atmosphere sets the origin of the column depth, $X=0$, and the effective nucleon interaction length $\lambda_N(E)$ is
\begin{equation}
	\lambda_N(E) = \frac{A}{N_A\sigma_{NA}(E)}\,
\end{equation}
for target mass number $A$ and Avogadro's number $N_A$. For nucleons,
\begin{equation}
	\phi_N(E,X)=\phi_N(E,0)\exp(-X/\Lambda_N),
	\quad \Lambda_N=\frac{\lambda_N}{1-Z_{NN}}\,.
\end{equation}
The $Z$-moment  for nucleon scattering is approximately $Z_{NN}\simeq
	0.23$~\cite{Bhattacharya:2015jpa}, which we assume is independent of energy.

\subsection{Stau flux from
	\texorpdfstring{$N\to\tilde{q}\to\stau$}{N->squark->stau}}

For staus, we consider directly the production and decay of squarks in the evaluation of $S_{N\to \stau} $. Since the decay length of staus is long and the decay length of squarks is very short, and since staus are unlikely to interact in the low density of the atmosphere, to a good approximation, only the source term contributes to the \stau\ flux at the surface of the Earth in eq. (\ref{eq:master}). It is our convention here to denote the flux of ${\stau}$ plus its antiparticle with $\phi_{\stau}$.
The cascade equation for the stau flux is therefore
\begin{equation}
	\frac{\partial\phi_\stau(E,X)}{\partial X}
	\simeq S_{N\to \stau}(E,X)\,,
\end{equation}
using the short hand of $S_{N\to \stau}\equiv S_{N\to \sqrk\to\stau}$.
The solution yields a nearly isotropic prompt stau flux at the Earth's surface.
Applying the same approximations to evaluate the $N\to\sqrk\to\stau$ contribution to the \stau\ flux that successfully approximate the prompt atmospheric neutrino flux \cite{Lipari:1993hd}, the stau flux can be written in terms of $Z$-moments as
\begin{equation}
	\label{eq:phi-stau-z}
	\phi_\stau(E)\simeq 2\frac{Z_{N\to\stau}(E)}{1-Z_{NN}(E)}\phi_N(E,X=0)\,.
\end{equation}
The factor of $2$ accounts for $\sqrk$-pair production, in which each \sqrk\ produces a \stau . Prompt squark decays in the atmosphere yield an isotropic stau flux in the high energy limit, assumed here.

\begin{figure}[htb]
	\centering
    \includegraphics[width=0.48\textwidth]{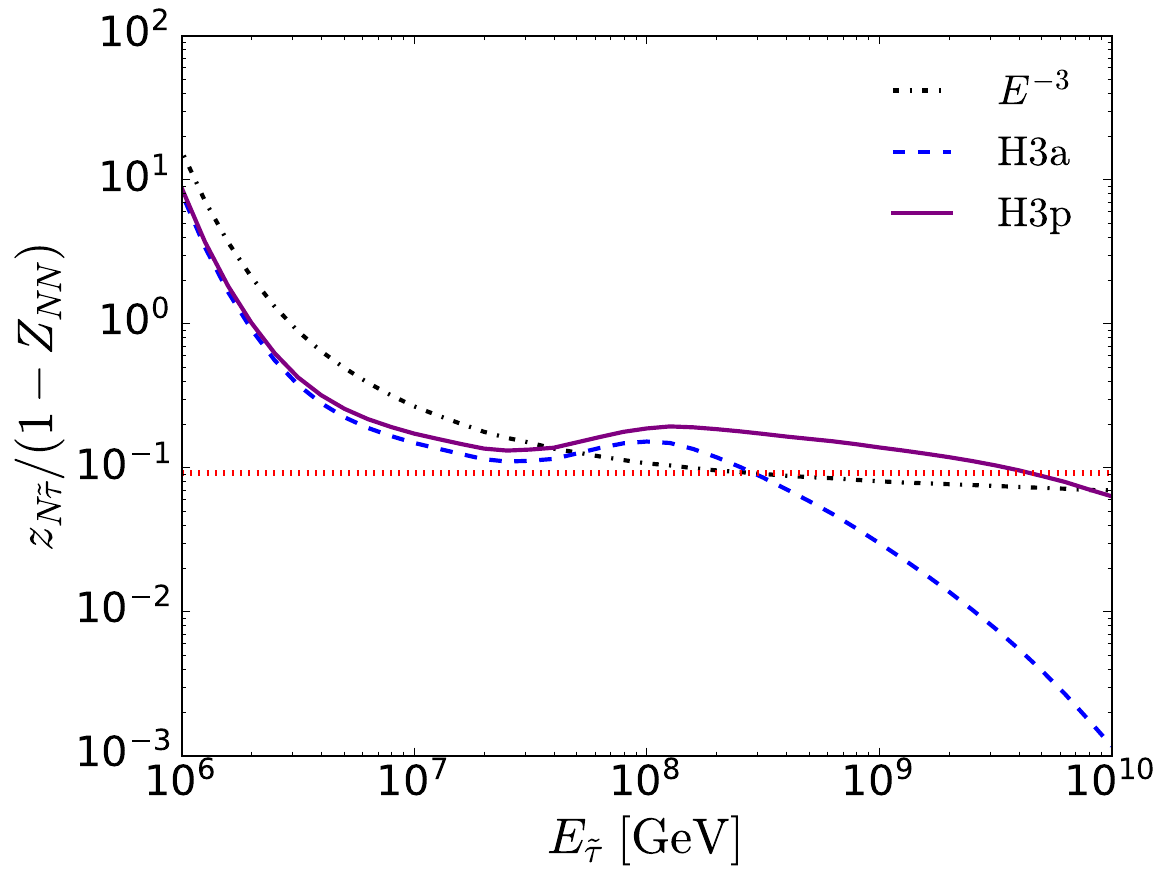}
    \includegraphics[width=0.48\textwidth]{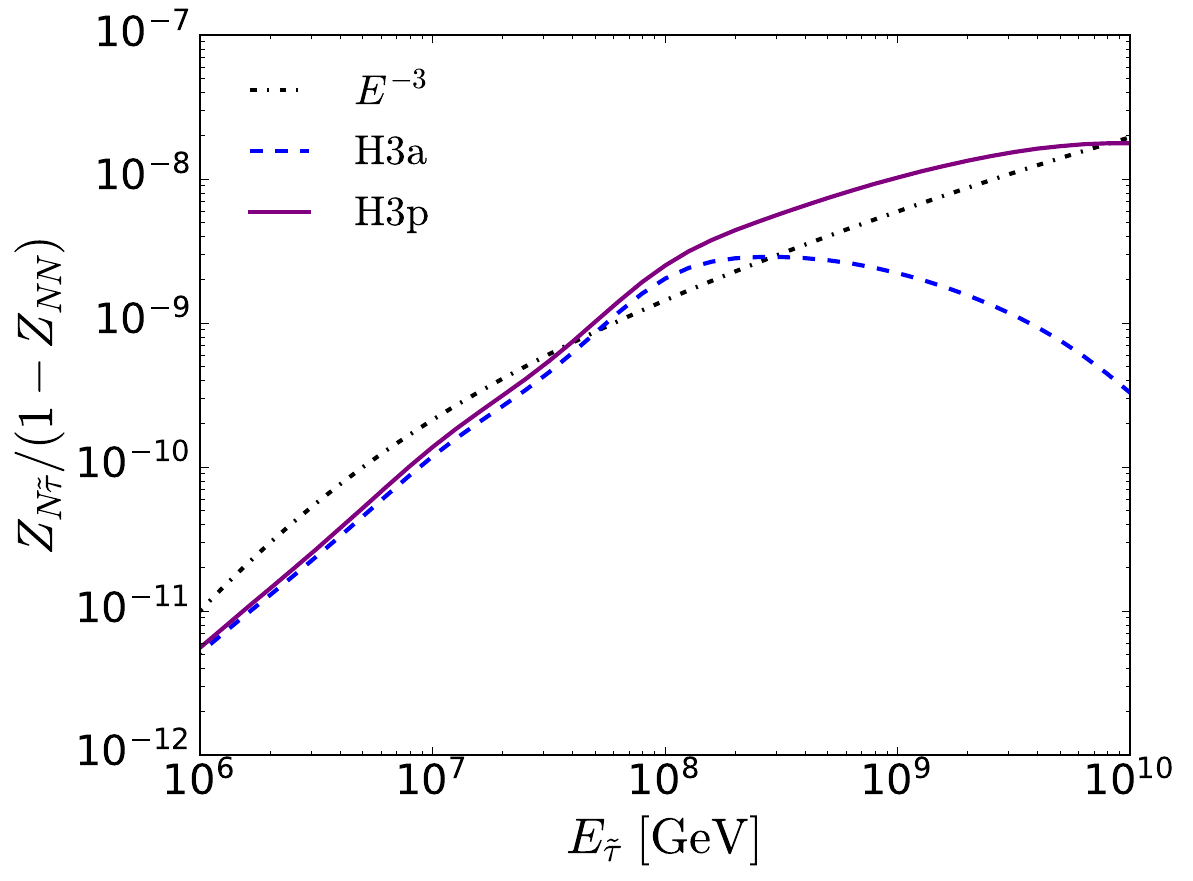}
	\caption{Left: The reduced $z$-moment defined in eq. (\ref{eq:zred}) multiplied by $(1-Z_{NN})^{-1}$ for cosmic ray nucleon production of \stau\ for for the H3a and H3p cosmic ray fluxes \cite{Gaisser:2011klf} and for an incident cosmic ray flux that scales as $E^{-3}$.
	The dotted red line shows the reduced $z$-moment using $d\sigma_{N\to \stau}/dx$ in eq. (\ref{eq:dsdx}), $\sigma_{p\to\stau}(E_N)\propto E_N^{0.58}$ and $\phi_N\propto E_N^{-3}$. Right: The moment $Z_{N\tilde{\tau}}$ defined by eq. (\ref{eq:Znstau}) scaled by $(1-Z_{NN})^{-1}$. In both panels, $\msq = 415$ GeV and $m_\stau=290$ GeV.}
	\label{fig:zstau}
\end{figure}

We use eq. (\ref{eq:dsdx-full}) and the $\sqrk$-pair production cross section to evaluate $Z_{N\stau}(E)$. 
The probability $P(E)$ of producing a \stau\ pair in cosmic ray--air collisions is
the ratio of the $NA$ \stau\ pair production cross section to the total cosmic ray--air cross section. We approximate $P(E)$ with the ratio of $pp$ cross sections,
\begin{equation}
	P(E)\simeq \frac{\sigma_{pp\to \stau}(E)}{\sigma_{pp}(E)}\,.
\end{equation}
It is convenient to define a new quantity, the reduced $z$-moment $z_{N \stau}$, such that
\begin{equation}
	Z_{N \stau}(E)\simeq \frac{\sigma_{NA\to \stau}(E)}{\sigma_{NA}(E)}
	z_{N \stau}(E)\simeq P(E) z_{N\stau}(E)\,.
    \label{eq:Znstau}
\end{equation}
This separates the production probability from the cosmic ray spectrum weighted stau energy distribution.
The reduced $z$-moment is therefore
\begin{equation}
	z_{N\stau}(E) =
	\int_E dE_N
	\frac{\phi_N(E_N)}{\phi_N(E)}
	\frac{\sigma_{N\to\tilde{\tau}} (E_N) }{\sigma_{N\to\tilde{\tau}} (E)} \frac{1}{E_N} \frac{dn}{dx}\, .
	\label{eq:zred}
\end{equation}
We can approximate the $Z$-moment for stau production assuming $p=N$ and that the differential cross section for $N\to \stau$ as a function of $x=E/E_N$ for $E=E_\stau$ scales with energy according to $\sigma_{N\to\stau}(E_N)$, namely that $(d\sigma_{N\to\stau}(x,E_N)/dx)/\sigma_{N\to\stau}(E_N)$ is independent of energy and $d\sigma_{N\to\stau}(x,E_N)/dx\sim \exp(-bx)$.

\begin{figure}[htb]
	\centering
    \includegraphics[width=0.55\textwidth]{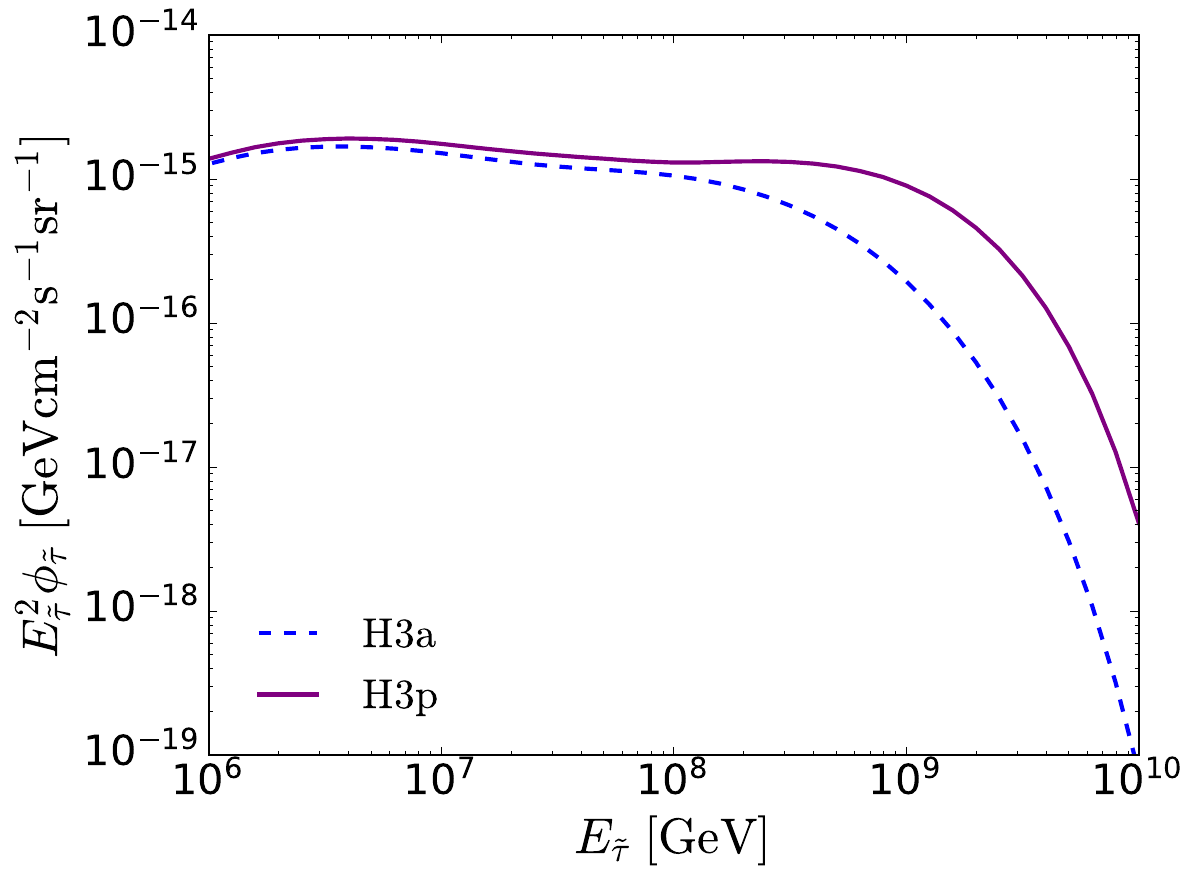}
	\caption{The atmospheric $\stau$ flux as a function of $\stau$ energy for H3p and H3a cosmic ray flux models for $\msq = 415$ GeV and $m_\stau=290$ GeV. For $m_\sqrk = 1.3$ TeV and $m_\stau = 490$ GeV, the atmospheric flux reduces by a factor of $\sim 1/330$ times the flux shown for $E_\stau=10^8$ GeV to $\sim 1/160$ for $E_\stau=10^{11}$ GeV.}
	\label{fig:zstau-flux}
\end{figure}

The  
scaled reduced $z$-moments $z_{N\stau}/(1-Z_{NN})$  and  and scaled $Z$-moments $Z_{N\stau}/(1-Z_{NN})$ for several different input cosmic ray fluxes and $\msq = 415$ GeV and $m_\stau=290$ GeV are shown in the left and right panels of figure \ref{fig:zstau}. The dot-dashed  line approximates the cosmic ray flux by
$\phi_N(E,X=0)\sim E^{-3}$. The dashed and solid curves in figure \ref{fig:zstau} show the evaluations of $z_{N\stau}/(1-Z_{NN})$ and $Z_{N\stau}/(1-Z_{NN})$ using Gaisser's models of the cosmic ray flux with 3 populations of cosmic rays following Hillas \cite{Gaisser:2011klf} with mixed (H3a) and only proton (H3p) components of the third population, respectively. Using the analytic approximation to the $d\sigma_{N\to \stau}/dx$ in eq. (\ref{eq:dsdx}) and also approximating $\sigma_{N\to\stau}(E_N)\propto E_N^{0.58}$
and $\phi_N\propto E_N^{-3}$, the value of $z_{N\stau}/(1-Z_{NN})\simeq 0.1$ is shown by the red dotted line in the left panel of  figure \ref{fig:zstau}. The ratio ${\sigma_{N\to\tilde{\tau}} (E_N) }/{\sigma_{N\to\tilde{\tau}} (E)}$ in eq. (\ref{eq:zred}) is responsible for the steep decrease in the reduced $z$-moment below $\sim 10^8$ GeV. The steep decrease in the reduced $z$-moment at high energy for the H3a cosmic ray flux comes from the steep fall in its cosmic ray energy per-nucleon distribution above $\sim 10^8$ GeV \cite{Gaisser:2011klf}. The middle panel of figure  \ref{fig:stau-xsec} shows that above $E_N\sim 10^8$, the cross sections for the two benchmark mass sets are proportional to each other, so for $E_\stau\gtrsim 10^8$ GeV, the reduced $z$-moments shown in the left panel of figure \ref{fig:zstau} are reliable for both benchmark mass sets.
The right panel shows that the full $Z$-moment increases as a function of energy, then for the H3a cosmic ray flux, it decreases. 
The prompt stau (particle plus antiparticle) flux is determined by eq. (\ref{eq:phi-stau-z}).
Figure \ref{fig:zstau-flux} shows the prompt stau flux for the default values of $\msq = 415$ GeV and $m_\stau=290$ GeV for H3p and H3a cosmic ray flux models. We will use the H3p cosmic ray flux as our default input flux. This will give us most optimistic results.  Use of other cosmic ray flux with heavier composition would result in lower event rates.   For $E_\stau\lesssim 10^8$ GeV, the cosmic ray flux model uncertainties are relatively small.

It is interesting to ask if secondary interactions from hadrons produced in $N$-air showers can also produce squarks that contribute to $\phi_\stau(E)$. 
The factor $z_{N\stau}(E)$ in eq. (\ref{eq:zred}) includes
$N\to\tilde{q}\to\stau$, however, multiple hadrons are produced in the CR--Air interaction, including pions that can re-interact and product squarks so that $N\to\pi^\pm\to\tilde{q}\to\stau$ contributes to the atmospheric \stau\ flux. Typically, the leading production process dominates the atmospheric flux prediction because of the convolution of the differential cross section for particle production with the steeply falling cosmic ray nucleon flux.
At the high energies considered here,
the cascade equations for $\pi^+$ and for $\pi^-$ are in the high energy limit where pion re-interaction dominates over pion decays, so
\begin{equation}
	\label{eq:pionflux}
	\frac{\partial \phi_\pi(E,X)}{\partial X}\simeq -\frac{\phi_\pi(E,X)}{\lambda_\pi(E)}
	+S_{N\to \pi}(E,X)+S_{\pi\to\pi}(E,X)\,,
\end{equation}
for pion interaction length $\lambda_
\pi$, with the solution
\begin{equation}
	\phi_\pi(E,X)=\phi_N(E,0)\frac{Z_{N\pi}}{1-Z_{NN}}\frac{\Lambda_\pi}{\Lambda_\pi-\Lambda_N}\biggl(\exp(-X/\Lambda_\pi)- \exp(-X/\Lambda_N)\biggr) \,,
\end{equation}
with $\Lambda_\pi=\lambda_\pi/(1-Z_{\pi\pi})$.
Including both cosmic ray nucleon and pion production of staus, the cascade equation for staus is approximately
\begin{equation}
	\frac{\partial\phi_{\stau}(E,X)}{\partial X}
	\simeq  S_{N\to \stau}(E,X)+S_{\pi\to\stau}(E,X)\,.
\end{equation}
The approximate solution for the sum of $\stau^-+\stau^+$ is
\begin{eqnarray}
	\label{phi-stau-z2}
	\phi_{\stau}(E)&\simeq& 2\Biggl[\frac{Z_{N\to\stau}(E)}{1-Z_{NN}(E)} +
	\frac{Z_{N\to\pi}(E)}{1-Z_{NN}(E)}
	\frac{Z_{\pi\to\stau}(E)}{1-Z_{\pi\pi}(E)}\Biggr]
	\phi_N(E,X=0)\\
	\nonumber
	&=& 2 \frac{Z_{N\to\stau}(E)}{1-Z_{NN}(E)}
	\Biggl[1+{Z_{N\to \pi}}\frac{1-Z_{NN}(E)}{1-Z_{\pi\pi}(E)}\frac{Z_{\pi\to\stau}(E)}{Z_{N\to\stau}(E)}
	\Biggr]\phi_N(E,X=0)\,.
\end{eqnarray}
The first term in the brackets reproduces eq. (\ref{eq:phi-stau-z}).

To estimate the contributions from pions to the atmospheric \stau\ flux, we use the spectrum-weighted $Z$-moments $Z_{p\pi}$ and $Z_{\pi\pi}$ for cosmic ray energies of 1 TeV from ref. \cite{Gaisser:2016uoy} for a cosmic ray flux of $E^{-2.7}$ evaluated using Sibyll 2.3:
\begin{eqnarray}
	\label{eq:zppip}
	Z_{p\to\pi^+}&=& 0.040\\
	\label{eq:zppim}
	Z_{p\to\pi^-}&=& 0.026\\
	\label{eq:zpipi}
	Z_{\pi^+\pi^+} &=&0.206\,.
\end{eqnarray}
We take $Z_{N\to\pi}=Z_{p\to \pi^+}+Z_{p\to\pi^-}$ since both $\pi^+$ and $\pi^-$ can produced \sqrk\ pairs, which we assume is independent of pion charge. We take
$Z_{\pi\pi}=Z_{\pi^+\pi^+}=Z_{\pi^-\pi^-}$.
The Sibyll 2.3 values \cite{Gaisser:2016uoy} are roughly consistent with the $Z$-moments in refs.~\cite{Gondolo:1995fq} and \cite{Fedynitch:2022vty}. The $Z$-moments are weakly energy dependent, so the energy independent approximation is sufficient here.
From eq.~\eqref{phi-stau-z2}, we can estimate the correction from
$N\to\pi^\pm\to\tilde{q}\to\stau$ to the flux of staus at Earth from
$N\to\tilde{q}\to\stau$. Using eqs.~\eqref{eq:zppip}-\eqref{eq:zpipi},
\begin{equation}
	\Biggl[1+{Z_{N\to \pi}}\frac{1-Z_{NN}(E)}{1-Z_{\pi\pi}(E)}\frac{Z_{\pi\to\stau}(E)}{Z_{N\to\stau}(E)}
	\Biggr]\simeq \Biggl[ 1+0.064\frac{Z_{\pi\to\stau}(E)}{Z_{N\to\stau}(E)}
		\Biggr]  \,.
\end{equation}
The $N-N$ and $\pi^\pm - N$ cross sections are comparable at high
energies, so we anticipate that the corresponding $Z$-moment ratio
${Z_{\pi\to\stau}(E)}/{Z_{N\to\stau}(E)}\sim 1$.
With
${Z_{\pi\to\stau}(E)}/{Z_{N\to\stau}(E)}\lesssim 1.5$, \stau\ production from
pion re-interactions in the atmosphere is $\lesssim 10\%$ of \stau\ production
by primary cosmic rays via $N\to\tilde{q}\to\stau$. Therefore, we conclude that for the relevant parameter choices for 
$m_\stau$ and $m_\sqrk$, the solid curve in figure \ref{fig:zstau-flux} is a reliable estimate of the stau flux at the Earth's surface.

\section{Stau propagation and detection}
\label{sec:stau-prop}
\subsection{Stau survival probability in Earth}

\begin{figure}
	\centering
	\begin{subfigure}{0.34\textwidth}
		\includegraphics[width=\textwidth]{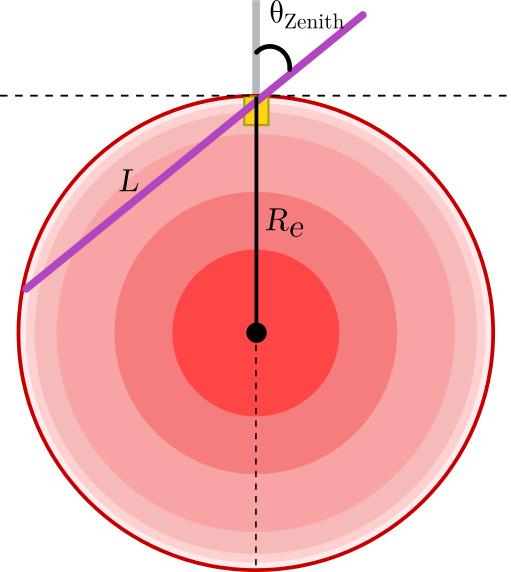}
	\end{subfigure}
	\begin{subfigure}{0.64\textwidth}
		\includegraphics[width=0.95\textwidth]{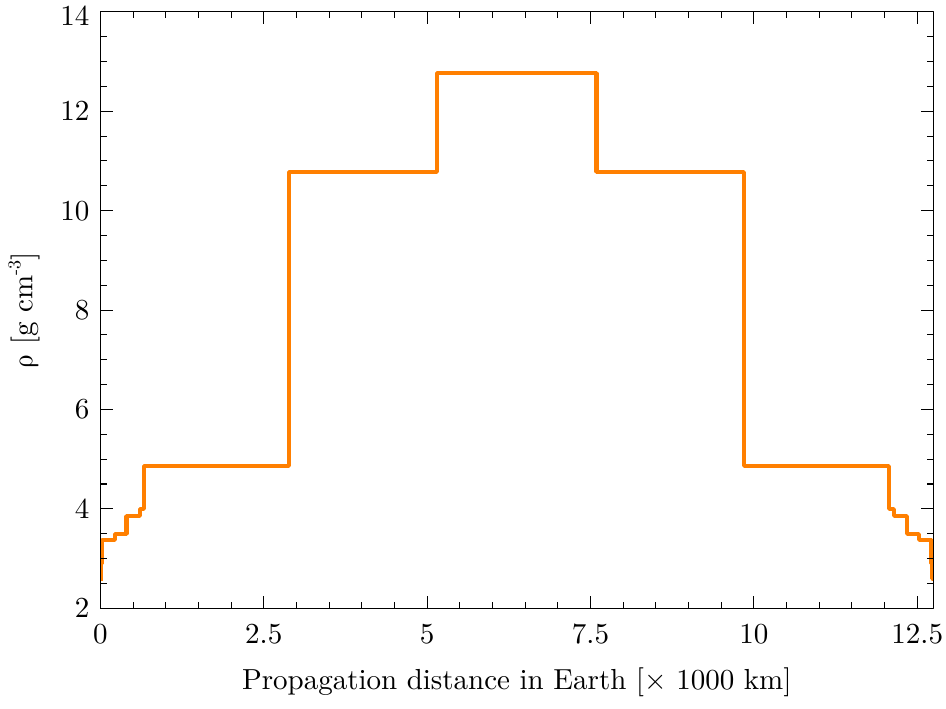}
	\end{subfigure}
	\caption{Left: Representation of a particle surviving a distance $L$ through Earth
		until reaching IceCube, at
		the zenith angle  $\theta_{\rm Zenith}$.\quad
		Right: Profile of the Earth's density as a function of distance traversed by a
		particle travelling along the Earth's diameter $\theta_{\rm Zenith}=0^\circ$.}
	\label{fig:earth}
\end{figure}

Given a \stau\ flux at the surface of the Earth, detection of \stau\ signals requires an understanding of the \stau\ survival probability as it transits the Earth and how its energy is degraded. A schematic of the \stau\ trajectory in the Earth is shown in the left panel of figure \ref{fig:earth}.

A \stau\ of mass $m_\stau$ loses energy as it traverses the distance d$\ell$ in
the Earth with density $\rho(\ell)$. The energy loss for very high energy  staus is  approximately described by \cite{Albuquerque:2003mi,Reno:2005si,Huang:2006ie,Anchordoqui:2019utb}:
\begin{subequations}
	\begin{equation}
		\frac{\diff E_\stau}{\diff \ell} = - b\left( m_\stau, E_\stau \right) \rho \left( \ell \right) E_\stau\,,
		\label{eq:elossmain}
	\end{equation}
 In standard rock ($\rho = 2.65$ g/cm$^3$), the energy loss parameter is approximately  \cite{Anchordoqui:2019utb}
	\begin{equation}
		b\left( m_\stau, E_\stau \right) = 4.8\times 10^{-10} \left( \frac{10^3\text{ GeV}}{m_\stau} \right)^{1.25}
		\left( 1 + 0.073 \log \frac{E_\stau}{10^9 \text{ GeV}} \right)\ \frac{\rm cm^2}{\rm g}\,.
		\label{eq:b}
	\end{equation}
	\label{eq:eloss}
\end{subequations}
As we show below, most of the \stau\ signal will come from angles up to $10^\circ$ below horizontal, where $\rho\lesssim 3.4$ g/cm$^3$, so the energy loss parameter in rock is a reasonable approximation.
The density
$\rho\left( \ell \right)$ is taken to be the PREM
profile~\cite{DZIEWONSKI1981297}
of the Earth's density (see figure \ref{fig:earth}, right panel).

The probability of the \stau\ surviving a distance $\ell$ is given by
\begin{subequations}
	\begin{equation}
		\frac{\diff \psurv}{\diff \ell} = -\frac{\psurv}{\ldec}
		\label{eq:psurv}
	\end{equation}
	\text{with the \stau\ decay length}
	\begin{equation}
		\ldec (m_\stau, \tau_\stau, E_\stau) = \frac{c\tau_\stau E_\stau}{m_\stau} \sqrt{1- \frac{m_\stau^2}{E_\stau^2}}
		\label{eq:ldec}
	\end{equation}
\end{subequations}
and $E_\stau (\ell)$ determined by eq.~\eqref{eq:eloss}.
To ensure that the stau is potentially detectable as a high-energy track at IceCube-like
detectors, and does not end up decaying either before reaching or exiting the
detector, we consider the case of large \stau\ lifetimes where $c\tau_\stau \geqslant 10$
m. The decay length is approximately
\begin{equation}
	\ldec (m_\stau, \tau_\stau, E_\stau) \simeq 3.4\times 10^3 \ {\rm km}\cdot \frac{c\tau_\stau }{10\ {\rm m}}\cdot \frac{290\ {\rm GeV}}{m_\stau} \cdot \frac{E_\stau}{10^6\ {\rm GeV}}\,.
\end{equation}
In the discussion below, we will be interested in $E_\stau \gtrsim 10^7-10^8$ GeV, so $\lambda_{\rm dec}$ is large compared to the radius of the Earth.
The long lifetime and relatively low electromagnetic energy loss of staus mean that the angular distribution of \stau\ events is different than the angular distribution of $\nu_\mu\to\mu$ events. However,
we note that the relatively long \stau\ lifetime makes the probability of \stau\ decays within detectable volumes very small, making them invisible to radio-detectors
like ANITA. Consequently, staus with $c\tau_\stau \geqslant 10$
m cannot account for the unusual events detected by ANITA coming from relatively large angles below the Earth's horizon
\cite{ANITA:2016vrp,ANITA:2018sgj}.

\subsection{Detecting \stau s as up-going tracks at IceCube and IceCube-Gen\textit{2}}

We now turn to the detection of staus by energy deposition along their trajectories in an IceCube-Gen\textit{2}-like detector.
The detection of \stau\  tracks in IceCube competes against the background
from muons produced from the atmospheric and diffuse extra-galactic muon neutrino flux.
As noted in eq. (\ref{eq:elossmain}), heavy \stau\ electromagnetic energy loss is significantly less than muon electromagnetic energy loss for equal initial energies. Long-lived high energy \stau\ signals can in principle appear at larger angles below the horizon than standard model $\nu_\mu\to \mu$ events. The reduced energy loss of staus, however, make them more difficult to detect.

Based on electromagnetic energy loss, \stau\ tracks could be interpreted as muons with a lower energy. The energy loss $\Delta E$ over a distance $\Delta \ell$ for a muon is $\Delta E\simeq b_\mu \rho E_\mu \Delta \ell $. The same energy loss $\Delta E$ for a \stau\ over distance $\Delta \ell$ is $\Delta E\simeq b_\stau \rho E_\stau \Delta \ell $. The effective \stau\ energy that would be determiend by electromagnetic energy loss measurements, interpreting the $\stau$ signal as coming from a muon, is therefore 
\begin{equation}
	{E_\stau}^{\rm eff} \simeq \frac{b_{\stau}}{b_\mu} E_\stau\,.
\end{equation}
Numerically, for the energies of interest, ${b_{\stau}}/{b_\mu}\sim 5\times 10^{-4}$.
While it may be possible to distinguish between muons of energy $E_\mu$ and staus of energy $E_\stau = b_\mu E_\mu/b_\stau$
based on characteristics of the showers, our analysis here is based on using
${E_\stau}^{\rm eff}$.

For our analysis, given that we expect the number of up-going \stau\  track events to be considerably lower
than up-going muon track events from the muon neutrino flux, we set a lower energy threshold
for the track energy in IceCube $E_\mu>E_{\rm min}$ and $E_\stau^{\rm eff}>E_{\rm min}$  for
$E_\text{min} = 100$ TeV, below which any prompt \stau\ flux limited by
current constraints on the SUSY spectrum would be swamped by a combination of
atmospheric leptonic background and diffuse extragalactic astrophysical
neutrino fluxes.
The choice of a $100$~TeV threshold ensures that the conventional atmospheric
background of muon neutrinos from pion and kaon decays is nearly completely
cut off, leaving only extra-galactic neutrino flux as the dominant source of
background for a prompt atmospheric $\stau$ flux.  In any case, we consider up-going events. 

For our benchmark, we take $m_\stau\ = 490$ GeV and $E_{\rm min}=100$ TeV.  For these parameters, we find 
$\order{10^{-4}}$ \stau\ events in 10 years of run-time from such tracks at
IceCube. 
For the lighter \stau\, for example when its mass is $290$~GeV, the number of events rises to $\sim 0.05$.
Clearly, a considerably larger instrumented volume is required to constrain
input parameters to models with \stau\ production in the atmosphere.

\Gen, an expanded version of IceCube with instrumented volume larger by up to a
factor of 100 at the highest energies, has been proposed as a successor of
IceCube~\cite{Clark:2021fkg}.
With its increased volume, \Gen\ will potentially be able to gather at least a
few events for \stau\ fluxes in the scenario with $m_\stau=290$ GeV and $m_\sqrk = 415$ GeV.
It, therefore, provides a reasonable potential for eventually discovering
\stau\ tracks over a period of 10--20 years, at least for \stau\ with lighter
masses.

\begin{figure}
	\centering
	\includegraphics[width=0.75\textwidth]{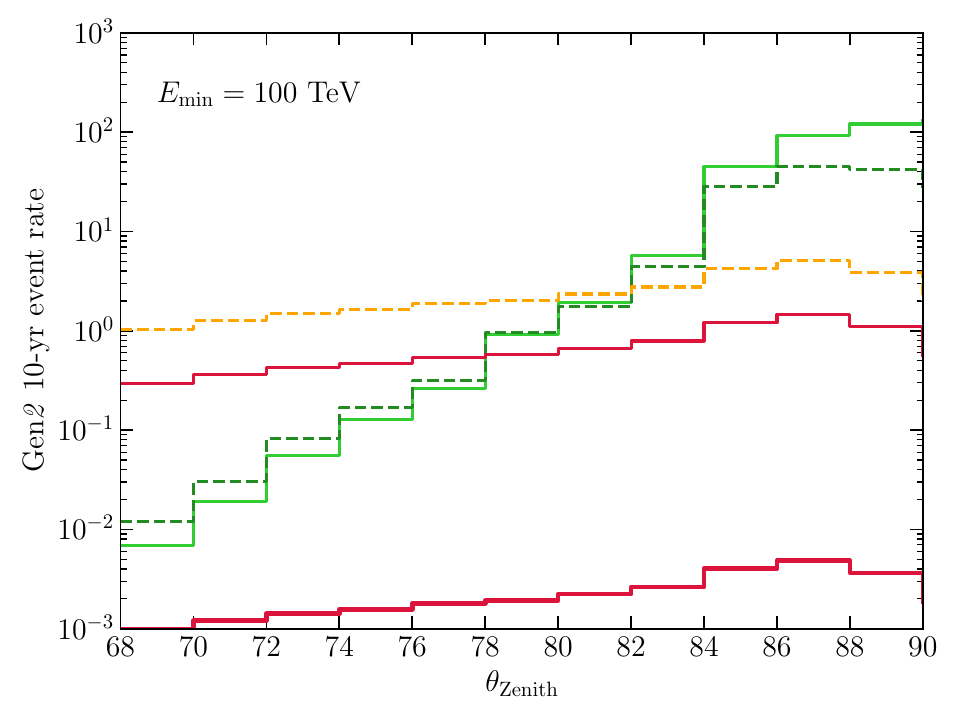}
	\caption{\label{fig:gen2events}Track event rates over 10 years at
		IceCube-Gen\textit{2} binned by the stau/muon track zenith angles
		at the detector for $(E_\mu,E_\stau^{\rm eff})$ muon equivalent energy
		thresholds of 100 TeV.
		The \stau\ tracks for $m_\stau = 290$  GeV for staus produced in the decay of squarks with $m_\sqrk = 415$ GeV and also tracks for $m_\stau = 490$ GeV from $m_\sqrk = 1.3$ TeV are represented
		by thin (upper) and thick (lower) solid red curves respectively.
		For comparison with Ref.~\cite{Meighen-Berger:2020eun}, we show results
		for the same analysis but with $m_\stau = 200$~GeV from
        $m_\sqrk = 385$ GeV (orange dashed curve). 
		Tracks from atmospheric and astrophysical muon neutrinos, which serve as
		the backgrounds to the detection of stau tracks are shown as dashed and
		solid green curves respectively.
		For the latter, we assume a best-fit power-law flux $d\Phi/dE \propto
			E^{-2.2} $ consistent with the IceCube best-fit obtained from the 10-yr
		through-going muon track analysis~\cite{IceCube:2021uhz}.
      }
\end{figure}

Overall, the total integrated flux of staus produced in the atmosphere leads
to track event rates in \Gen\ that are a few orders of magnitude lower than
those from charged-current interactions of atmospheric muon neutrinos combined with the muon neutrino flux
from astrophysical sources modeled with the  best-fit power-law \cite{IceCube:2021uhz}.
While the fluxes integrated over solid angle make \stau\ discovery challenging,
the low energy loss of staus mean that the angular distribution of the \stau\
flux that leads to tracks in detectors is different that the angular
distribution of muons that come from $\nu_\mu$ charged-current events that
occur outside of the detector and lead to
through-going muons.
As a result the rate of muon track events as a function of zenith angle
$\theta_{\rm Zenith}$ (see figure \ref{fig:earth}), whether from muon neutrinos
produced in the atmosphere or in astrophysical sources, steeply drops as one
looks for tracks from increasingly deeper below the horizon ($\theta_{\rm
		Zenith}< 90^\circ$).
In contrast, the \stau\ track-rate at equivalent energies remains relatively
flat as a function of the angle below the horizon.
Although the corresponding event rates are still low (a couple of events in 10 years
for $E_{\rm min}=100$ TeV even with $m_\stau = 290$~GeV for $68^\circ \lesssim
	\thzen\ \lesssim 70^\circ$),
looking at up-going tracks with $\thzen \leqslant 70 ^\circ$ provides a
potential for discovery for these events over and above the background.
We show 10-yr event rates for tracks from staus as well as $\nu_\mu$ from
atmospheric and astrophysical neutrino fluxes in figure~\figref{gen2events}
for $E_{\rm min}=100$ TeV.

In ref.~\cite{Meighen-Berger:2020eun}, the authors have looked at IceCube event
rates for low \stau\ masses of 100--200 GeV using MCEq modified to include the
generation of \stau\ particles.
To compare the number of events from similarly low \stau\ mass, we also show
stau event rates with $m_\stau=200$ GeV, evaluated with the $Z$-moment method.
The dashed green histograms in the left and right plots in
figure~\ref{fig:gen2events} represent the predicted number of events per
angular bin per decade for $m_\stau=200$ GeV.
The lower mass means the event rate increases by a factor of $\sim 3-4$,
however, it is difficult to reconcile such a small \stau\ mass with LHC limits
on \stau\ pair production. The event rates calculated here for $m_\stau=200$ GeV with $m_\sqrk = 385$ GeV, as dictated by the GMSB SPS 7 model, are significantly lower than the
rates of ref. \cite{Meighen-Berger:2020eun}.

\section{Conclusions}
\label{sec:conclusions}

We have considered production of long-lived staus in the interaction of cosmic rays with atmosphere.  Direct production of staus with mass above the current LHC limit of about $430-490$ GeV is suppressed due to the large stau mass which gives a very small cross section.  In a GMSB scenario, staus are also produced via the decay of squarks.  We have studied  TeV mass squarks,  produced in the collision of cosmic rays with 
atmospheric nuclei, which promptly decay into staus.  We have found that the track event rates at \Gen\ are about two orders of magnitude higher than in the case of direct staus production, however, overall, stau event rates are low.  We emphasize that calculation needs to include stau energy distribution, and we accomplish this by solving the coupled cascade equations analogous to the way we calculated the standard model prompt neutrinos.  We find that the charged track events from staus have weak dependence on the zenith angle because of the stau's long lifetime and relatively low electromagnetic energy loss, while events rates from background atmospheric and astrophysical neutrinos have steep fall for deeper angles. To compare with results in the literature \cite{Meighen-Berger:2020eun}, we show our results for  $m_\stau =290$ GeV,  in which case the stau signal becomes 
dominant for angles below 78 degrees, and is about factor of 3-4 lower than results shown in ref. \cite{Meighen-Berger:2020eun}.
Since null results from the LHC constrain the stau
mass in GMSB models to be larger than  490 GeV, detectors with
significantly larger instrumented volume than the planned
\Gen\ would be needed for their detection.

We have presented our results using H3p cosmic ray flux, which is higher than H3a flux by about $25\%$ at $\stau$ energies around $10^8$ GeV, energies  corresponding to $E_\stau^{\rm eff}$ of about  $100$ TeV.  This uncertainty due to the choice of cosmic ray flux becomes larger for higher energies, and projected rates decrease if one assumes that the high energy cosmic ray  primary particles are heavy nuclei.

Our analysis follows  the standard $Z$-moment based 
treatment of prompt atmospheric neutrino production, with
appropriate modifications to the relevant energy scales and
cross-sections.
This serves as a template for the study of production of
prompt BSM particles from cosmic-ray interactions in the
atmosphere more generally, for models where the final quasi-stable or stable particle 
originates from the decay of a
short-lived intermediary, and of their subsequent detection at
neutrino telescopes like IceCube and KM3NeT at present and
\Gen\ in the future.

\vskip 0.2in
\noindent{\bf Acknowledgments}
\vskip 0.2in

We thank Tyce DeYoung for helpful discussions.
This work is supported in part by the U.S.~Department of Energy
Grant DE-SC-0010113 (MHR) and in part by the U.S.~Department of Energy Grant
DE-FG02-13ER41976/DE-SC-0009913 (IS).

\bibliographystyle{JHEP}
\bibliography{stau}

\providecommand{\href}[2]{#2}\begingroup\raggedright\begin{thebibliography}{10}

\bibitem{Barr:2004br}
G.~D. Barr, T.~K. Gaisser, P.~Lipari, S.~Robbins and T.~Stanev, \emph{{A Three
  - dimensional calculation of atmospheric neutrinos}},
  \href{http://dx.doi.org/10.1103/PhysRevD.70.023006}{\emph{Phys. Rev. D} {\bf
  70} (2004) 023006}, [\href{http://arxiv.org/abs/astro-ph/0403630}{{\tt
  astro-ph/0403630}}].

\bibitem{Honda:2006qj}
M.~Honda, T.~Kajita, K.~Kasahara, S.~Midorikawa and T.~Sanuki,
  \emph{{Calculation of atmospheric neutrino flux using the interaction model
  calibrated with atmospheric muon data}},
  \href{http://dx.doi.org/10.1103/PhysRevD.75.043006}{\emph{Phys. Rev. D} {\bf
  75} (2007) 043006}, [\href{http://arxiv.org/abs/astro-ph/0611418}{{\tt
  astro-ph/0611418}}].

\bibitem{Sinegovskaya:2014pia}
T.~S. Sinegovskaya, A.~D. Morozova and S.~I. Sinegovsky, \emph{{High-energy
  neutrino fluxes and flavor ratio in the Earth\textquoteright{}s atmosphere}},
  \href{http://dx.doi.org/10.1103/PhysRevD.91.063011}{\emph{Phys. Rev. D} {\bf
  91} (2015) 063011}, [\href{http://arxiv.org/abs/1407.3591}{{\tt 1407.3591}}].

\bibitem{Gaisser:2002jj}
T.~K. Gaisser and M.~Honda, \emph{{Flux of atmospheric neutrinos}},
  \href{http://dx.doi.org/10.1146/annurev.nucl.52.050102.090645}{\emph{Ann.
  Rev. Nucl. Part. Sci.} {\bf 52} (2002) 153--199},
  [\href{http://arxiv.org/abs/hep-ph/0203272}{{\tt hep-ph/0203272}}].

\bibitem{Lipari:1993hd}
P.~Lipari, \emph{{Lepton spectra in the earth's atmosphere}},
  \href{http://dx.doi.org/10.1016/0927-6505(93)90022-6}{\emph{Astropart. Phys.}
  {\bf 1} (1993) 195--227}.

\bibitem{Pasquali:1998xf}
L.~Pasquali and M.~H. Reno, \emph{{Tau-neutrino fluxes from atmospheric
  charm}}, \href{http://dx.doi.org/10.1103/PhysRevD.59.093003}{\emph{Phys. Rev.
  D} {\bf 59} (1999) 093003}, [\href{http://arxiv.org/abs/hep-ph/9811268}{{\tt
  hep-ph/9811268}}].

\bibitem{Enberg:2008te}
R.~Enberg, M.~H. Reno and I.~Sarcevic, \emph{{Prompt neutrino fluxes from
  atmospheric charm}},
  \href{http://dx.doi.org/10.1103/PhysRevD.78.043005}{\emph{Phys. Rev. D} {\bf
  78} (2008) 043005}, [\href{http://arxiv.org/abs/0806.0418}{{\tt 0806.0418}}].

\bibitem{Gauld:2015kvh}
R.~Gauld, J.~Rojo, L.~Rottoli, S.~Sarkar and J.~Talbert, \emph{{The prompt
  atmospheric neutrino flux in the light of LHCb}},
  \href{http://dx.doi.org/10.1007/JHEP02(2016)130}{\emph{JHEP} {\bf 02} (2016)
  130}, [\href{http://arxiv.org/abs/1511.06346}{{\tt 1511.06346}}].

\bibitem{Bhattacharya:2015jpa}
A.~Bhattacharya, R.~Enberg, M.~H. Reno, I.~Sarcevic and A.~Stasto,
  \emph{{Perturbative charm production and the prompt atmospheric neutrino flux
  in light of RHIC and LHC}},
  \href{http://dx.doi.org/10.1007/JHEP06(2015)110}{\emph{JHEP} {\bf 06} (2015)
  110}, [\href{http://arxiv.org/abs/1502.01076}{{\tt 1502.01076}}].

\bibitem{Bhattacharya:2016jce}
A.~Bhattacharya, R.~Enberg, Y.~S. Jeong, C.~S. Kim, M.~H. Reno, I.~Sarcevic
  et~al., \emph{{Prompt atmospheric neutrino fluxes: perturbative QCD models
  and nuclear effects}},
  \href{http://dx.doi.org/10.1007/JHEP11(2016)167}{\emph{JHEP} {\bf 11} (2016)
  167}, [\href{http://arxiv.org/abs/1607.00193}{{\tt 1607.00193}}].

\bibitem{Zenaiev:2019ktw}
{\scshape PROSA} collaboration, O.~Zenaiev, M.~V. Garzelli, K.~Lipka, S.~O.
  Moch, A.~Cooper-Sarkar, F.~Olness et~al., \emph{{Improved constraints on
  parton distributions using LHCb, ALICE and HERA heavy-flavour measurements
  and implications for the predictions for prompt atmospheric-neutrino
  fluxes}}, \href{http://dx.doi.org/10.1007/JHEP04(2020)118}{\emph{JHEP} {\bf
  04} (2020) 118}, [\href{http://arxiv.org/abs/1911.13164}{{\tt 1911.13164}}].

\bibitem{Fedynitch:2022vty}
A.~Fedynitch and M.~Huber, \emph{{Data-driven hadronic interaction model for
  atmospheric lepton flux calculations}},
  \href{http://arxiv.org/abs/2205.14766}{{\tt 2205.14766}}.

\bibitem{Super-Kamiokande:2015qek}
{\scshape Super-Kamiokande} collaboration, E.~Richard et~al.,
  \emph{{Measurements of the atmospheric neutrino flux by Super-Kamiokande:
  energy spectra, geomagnetic effects, and solar modulation}},
  \href{http://dx.doi.org/10.1103/PhysRevD.94.052001}{\emph{Phys. Rev. D} {\bf
  94} (2016) 052001}, [\href{http://arxiv.org/abs/1510.08127}{{\tt
  1510.08127}}].

\bibitem{IceCube:2015mgt}
{\scshape IceCube} collaboration, M.~G. Aartsen et~al., \emph{{Measurement of
  the Atmospheric $\nu_e$ Spectrum with IceCube}},
  \href{http://dx.doi.org/10.1103/PhysRevD.91.122004}{\emph{Phys. Rev. D} {\bf
  91} (2015) 122004}, [\href{http://arxiv.org/abs/1504.03753}{{\tt
  1504.03753}}].

\bibitem{IceCube:2017cyo}
{\scshape IceCube} collaboration, M.~G. Aartsen et~al., \emph{{Measurement of
  the $\nu _{\mu}$ energy spectrum with IceCube-79}},
  \href{http://dx.doi.org/10.1140/epjc/s10052-017-5261-3}{\emph{Eur. Phys. J.
  C} {\bf 77} (2017) 692}, [\href{http://arxiv.org/abs/1705.07780}{{\tt
  1705.07780}}].

\bibitem{ANTARES:2021cwc}
{\scshape ANTARES} collaboration, A.~Albert et~al., \emph{{Measurement of the
  atmospheric $\nu_e$ and $\nu_\mu$ energy spectra with the ANTARES neutrino
  telescope}},
  \href{http://dx.doi.org/10.1016/j.physletb.2021.136228}{\emph{Phys. Lett. B}
  {\bf 816} (2021) 136228}, [\href{http://arxiv.org/abs/2101.12170}{{\tt
  2101.12170}}].

\bibitem{Kochanov:2021hkj}
A.~A. Kochanov, A.~D. Morozova, T.~S. Sinegovskaya and S.~I. Sinegovsky,
  \emph{{High-energy spectra of the atmospheric neutrinos: predictions and
  measurements}},  \href{http://arxiv.org/abs/2109.13000}{{\tt 2109.13000}}.

\bibitem{TelescopeArray:2018bya}
{\scshape Telescope Array} collaboration, R.~U. Abbasi et~al., \emph{{The
  Cosmic-Ray Energy Spectrum between 2 PeV and 2 EeV Observed with the TALE
  detector in monocular mode}},
  \href{http://dx.doi.org/10.3847/1538-4357/aada05}{\emph{Astrophys. J.} {\bf
  865} (2018) 74}, [\href{http://arxiv.org/abs/1803.01288}{{\tt 1803.01288}}].

\bibitem{PierreAuger:2020qqz}
{\scshape Pierre Auger} collaboration, A.~Aab et~al., \emph{{Measurement of the
  cosmic-ray energy spectrum above $2.5{\times} 10^{18}$ eV using the Pierre
  Auger Observatory}},
  \href{http://dx.doi.org/10.1103/PhysRevD.102.062005}{\emph{Phys. Rev. D} {\bf
  102} (2020) 062005}, [\href{http://arxiv.org/abs/2008.06486}{{\tt
  2008.06486}}].

\bibitem{Albuquerque:2003mi}
I.~Albuquerque, G.~Burdman and Z.~Chacko, \emph{{Neutrino telescopes as a
  direct probe of supersymmetry breaking}},
  \href{http://dx.doi.org/10.1103/PhysRevLett.92.221802}{\emph{Phys. Rev.
  Lett.} {\bf 92} (2004) 221802},
  [\href{http://arxiv.org/abs/hep-ph/0312197}{{\tt hep-ph/0312197}}].

\bibitem{Albuquerque:2006am}
I.~F.~M. Albuquerque, G.~Burdman and Z.~Chacko, \emph{{Direct detection of
  supersymmetric particles in neutrino telescopes}},
  \href{http://dx.doi.org/10.1103/PhysRevD.75.035006}{\emph{Phys. Rev. D} {\bf
  75} (2007) 035006}, [\href{http://arxiv.org/abs/hep-ph/0605120}{{\tt
  hep-ph/0605120}}].

\bibitem{Reno:2005si}
M.~H. Reno, I.~Sarcevic and S.~Su, \emph{{Propagation of supersymmetric charged
  sleptons at high energies}},
  \href{http://dx.doi.org/10.1016/j.astropartphys.2005.06.002}{\emph{Astropart.
  Phys.} {\bf 24} (2005) 107--115},
  [\href{http://arxiv.org/abs/hep-ph/0503030}{{\tt hep-ph/0503030}}].

\bibitem{Huang:2006ie}
Y.~Huang, M.~H. Reno, I.~Sarcevic and J.~Uscinski, \emph{{Weak interactions of
  supersymmetric staus at high energies}},
  \href{http://dx.doi.org/10.1103/PhysRevD.74.115009}{\emph{Phys. Rev.} {\bf
  D74} (2006) 115009}, [\href{http://arxiv.org/abs/hep-ph/0607216}{{\tt
  hep-ph/0607216}}].

\bibitem{Ahlers:2006pf}
M.~Ahlers, J.~Kersten and A.~Ringwald, \emph{{Long-lived staus at neutrino
  telescopes}},
  \href{http://dx.doi.org/10.1088/1475-7516/2006/07/005}{\emph{JCAP} {\bf 07}
  (2006) 005}, [\href{http://arxiv.org/abs/hep-ph/0604188}{{\tt
  hep-ph/0604188}}].

\bibitem{Ahlers:2007js}
M.~Ahlers, J.~I. Illana, M.~Masip and D.~Meloni, \emph{{Long-lived staus from
  cosmic rays}},
  \href{http://dx.doi.org/10.1088/1475-7516/2007/08/008}{\emph{JCAP} {\bf 08}
  (2007) 008}, [\href{http://arxiv.org/abs/0705.3782}{{\tt 0705.3782}}].

\bibitem{Ando:2007ds}
S.~Ando, J.~F. Beacom, S.~Profumo and D.~Rainwater, \emph{{Probing new physics
  with long-lived charged particles produced by atmospheric and astrophysical
  neutrinos}},
  \href{http://dx.doi.org/10.1088/1475-7516/2008/04/029}{\emph{JCAP} {\bf 04}
  (2008) 029}, [\href{http://arxiv.org/abs/0711.2908}{{\tt 0711.2908}}].

\bibitem{Meighen-Berger:2020eun}
S.~Meighen-Berger, M.~Agostini, A.~Ibarra, K.~Krings, H.~Niederhausen,
  A.~Rappelt et~al., \emph{{New constraints on supersymmetry using neutrino
  telescopes}},
  \href{http://dx.doi.org/10.1016/j.physletb.2020.135929}{\emph{Phys. Lett. B}
  {\bf 811} (2020) 135929}, [\href{http://arxiv.org/abs/2005.07523}{{\tt
  2005.07523}}].

\bibitem{CMS:2012wcg}
{\scshape CMS} collaboration, S.~Chatrchyan et~al., \emph{{Search for heavy
  long-lived charged particles in $pp$ collisions at $\sqrt{s}=7$ TeV}},
  \href{http://dx.doi.org/10.1016/j.physletb.2012.06.023}{\emph{Phys. Lett. B}
  {\bf 713} (2012) 408--433}, [\href{http://arxiv.org/abs/1205.0272}{{\tt
  1205.0272}}].

\bibitem{ATLAS:2012urj}
{\scshape ATLAS} collaboration, G.~Aad et~al., \emph{{Searches for heavy
  long-lived sleptons and R-Hadrons with the ATLAS detector in $pp$ collisions
  at $\sqrt{s}=7$ TeV}},
  \href{http://dx.doi.org/10.1016/j.physletb.2013.02.015}{\emph{Phys. Lett. B}
  {\bf 720} (2013) 277--308}, [\href{http://arxiv.org/abs/1211.1597}{{\tt
  1211.1597}}].

\bibitem{ATLAS:2014fka}
{\scshape ATLAS} collaboration, G.~Aad et~al., \emph{{Searches for heavy
  long-lived charged particles with the ATLAS detector in proton-proton
  collisions at $ \sqrt{s}=8 $ TeV}},
  \href{http://dx.doi.org/10.1007/JHEP01(2015)068}{\emph{JHEP} {\bf 01} (2015)
  068}, [\href{http://arxiv.org/abs/1411.6795}{{\tt 1411.6795}}].

\bibitem{ATLAS:2019gqq}
{\scshape ATLAS} collaboration, M.~Aaboud et~al., \emph{{Search for heavy
  charged long-lived particles in the ATLAS detector in 36.1 fb$^{-1}$ of
  proton-proton collision data at $\sqrt{s} = 13$ TeV}},
  \href{http://dx.doi.org/10.1103/PhysRevD.99.092007}{\emph{Phys. Rev. D} {\bf
  99} (2019) 092007}, [\href{http://arxiv.org/abs/1902.01636}{{\tt
  1902.01636}}].

\bibitem{CMS:2016kce}
{\scshape CMS} collaboration, V.~Khachatryan et~al., \emph{{Search for
  long-lived charged particles in proton-proton collisions at $\sqrt s=$ 13
  TeV}}, \href{http://dx.doi.org/10.1103/PhysRevD.94.112004}{\emph{Phys. Rev.
  D} {\bf 94} (2016) 112004}, [\href{http://arxiv.org/abs/1609.08382}{{\tt
  1609.08382}}].

\bibitem{Giudice:1998bp}
G.~Giudice and R.~Rattazzi, \emph{{Theories with gauge mediated supersymmetry
  breaking}},
  \href{http://dx.doi.org/10.1016/S0370-1573(99)00042-3}{\emph{Phys. Rept.}
  {\bf 322} (1999) 419--499}, [\href{http://arxiv.org/abs/hep-ph/9801271}{{\tt
  hep-ph/9801271}}].

\bibitem{Allanach:2002nj}
B.~C. Allanach et~al., \emph{{The Snowmass Points and Slopes: Benchmarks for
  SUSY Searches}},
  \href{http://dx.doi.org/10.1007/s10052-002-0949-3}{\emph{Eur. Phys. J. C}
  {\bf 25} (2002) 113--123}, [\href{http://arxiv.org/abs/hep-ph/0202233}{{\tt
  hep-ph/0202233}}].

\bibitem{Heisig:2011dr}
J.~Heisig and J.~Kersten, \emph{{Production of long-lived staus in the
  Drell-Yan process}},
  \href{http://dx.doi.org/10.1103/PhysRevD.84.115009}{\emph{Phys. Rev. D} {\bf
  84} (2011) 115009}, [\href{http://arxiv.org/abs/1106.0764}{{\tt 1106.0764}}].

\bibitem{Feng:2015wqa}
J.~L. Feng, S.~Iwamoto, Y.~Shadmi and S.~Tarem, \emph{{Long-Lived Sleptons at
  the LHC and a 100 TeV Proton Collider}},
  \href{http://dx.doi.org/10.1007/JHEP12(2015)166}{\emph{JHEP} {\bf 12} (2015)
  166}, [\href{http://arxiv.org/abs/1505.02996}{{\tt 1505.02996}}].

\bibitem{Gaisser:2011klf}
T.~K. Gaisser, \emph{{Spectrum of cosmic-ray nucleons, kaon production, and the
  atmospheric muon charge ratio}},
  \href{http://dx.doi.org/10.1016/j.astropartphys.2012.02.010}{\emph{Astropart.
  Phys.} {\bf 35} (2012) 801--806}, [\href{http://arxiv.org/abs/1111.6675}{{\tt
  1111.6675}}].

\bibitem{Gaisser:2016uoy}
T.~K. Gaisser, R.~Engel and E.~Resconi, \emph{{Cosmic Rays and Particle
  Physics}: {2nd Edition}}.
\newblock Cambridge University Press, 6, 2016.

\bibitem{Gondolo:1995fq}
P.~Gondolo, G.~Ingelman and M.~Thunman, \emph{{Charm production and high-energy
  atmospheric muon and neutrino fluxes}},
  \href{http://dx.doi.org/10.1016/0927-6505(96)00033-3}{\emph{Astropart. Phys.}
  {\bf 5} (1996) 309--332}, [\href{http://arxiv.org/abs/hep-ph/9505417}{{\tt
  hep-ph/9505417}}].

\bibitem{Anchordoqui:2019utb}
L.~A. Anchordoqui et~al., \emph{{The pros and cons of beyond standard model
  interpretations of ANITA events}},
  \href{http://dx.doi.org/10.22323/1.358.0884}{\emph{PoS} {\bf ICRC2019} (2020)
  884}, [\href{http://arxiv.org/abs/1907.06308}{{\tt 1907.06308}}].

\bibitem{DZIEWONSKI1981297}
A.~M. Dziewonski and D.~L. Anderson, \emph{Preliminary reference earth model},
  {\emph{Physics of the Earth and Planetary Interiors} {\bf 25} (1981) 297 --
  356}.

\bibitem{ANITA:2016vrp}
{\scshape ANITA} collaboration, P.~W. Gorham et~al., \emph{{Characteristics of
  Four Upward-pointing Cosmic-ray-like Events Observed with ANITA}},
  \href{http://dx.doi.org/10.1103/PhysRevLett.117.071101}{\emph{Phys. Rev.
  Lett.} {\bf 117} (2016) 071101}, [\href{http://arxiv.org/abs/1603.05218}{{\tt
  1603.05218}}].

\bibitem{ANITA:2018sgj}
{\scshape ANITA} collaboration, P.~W. Gorham et~al., \emph{{Observation of an
  Unusual Upward-going Cosmic-ray-like Event in the Third Flight of ANITA}},
  \href{http://dx.doi.org/10.1103/PhysRevLett.121.161102}{\emph{Phys. Rev.
  Lett.} {\bf 121} (2018) 161102}, [\href{http://arxiv.org/abs/1803.05088}{{\tt
  1803.05088}}].

\bibitem{Clark:2021fkg}
{\scshape IceCube-Gen2} collaboration, B.~Clark, \emph{{The IceCube-Gen2
  Neutrino Observatory}},
  \href{http://dx.doi.org/10.1088/1748-0221/16/10/C10007}{\emph{JINST} {\bf 16}
  (2021) C10007}, [\href{http://arxiv.org/abs/2108.05292}{{\tt 2108.05292}}].

\bibitem{IceCube:2021uhz}
{\scshape IceCube} collaboration, R.~Abbasi et~al., \emph{{Improved
  Characterization of the Astrophysical Muon\textendash{}neutrino Flux with 9.5
  Years of IceCube Data}},
  \href{http://dx.doi.org/10.3847/1538-4357/ac4d29}{\emph{Astrophys. J.} {\bf
  928} (2022) 50}, [\href{http://arxiv.org/abs/2111.10299}{{\tt 2111.10299}}].

\end{thebibliography}\endgroup
\end{document}